\documentclass[Journal,10pt]{IEEEtran}
\newcommand{\figref}{Fig.~\ref}

\usepackage{color}
\usepackage{placeins}
\usepackage{float}
\usepackage{tabularx,colortbl}
\usepackage{setspace}
\usepackage{cite,graphicx,amsmath,psfrag,bm}
\usepackage{subfigure}
\usepackage[usenames,dvipsnames]{xcolor}
\usepackage{mathabx}
\usepackage{cancel}
\usepackage{epstopdf}
\usepackage{pstool}
\usepackage{caption}

\begin{document}

\title{Loss-Gain Equalized Reconfigurable Phaser for\\ Dynamic Radio Analog Signal Processing (R-ASP) }

\author{Lianfeng Zou,~\IEEEmembership{Student member,~IEEE,} Shulabh Gupta,~\IEEEmembership{member,~IEEE,}
        Christophe Caloz,~\IEEEmembership{Fellow,~IEEE}
\thanks{L. Zou, S. Gupta and C. Caloz are with the Department
of Electrical Engineering and Poly-Grames research center, \'{E}cole Polytechnique de Montr\'{e}al, Montr\'{e}al, Qu\'{e}bec, H3T 1J4, Canada (e-mail: lianfeng.zou@polymtl.ca). C. Caloz is also with King Abdulaziz University, Jeddah, Saudi Arabia (email: christophe.caloz@polymtl.ca).}}


\maketitle

\begin{abstract}
We present a loss-gain equalized reconfigurable phaser for dynamic radio analog signal processing (R-ASP). Such a phaser provides real-time tunable group delay response with all-pass transmission. We propose a lumped loss-gain implementation, where tuning and equalization are mostly easily achieved. A theoretical study derives the transfer function and the fundamental characteristics of the device. The phaser is finally experimentally demonstrated, first using a single loss-gain pair and finally a three cascaded loss-gain pair structure with full reconfigurability , where up-chirp and down-chirp group delays are shown for illustration. It is expected that this phaser will find wide applications in radio analog signal processing (R-ASP) systems requiring dynamic adaptability.
\end{abstract}


\IEEEpeerreviewmaketitle
\section{Introduction}\label{SEC:INTRO}

Radio-analog signal processing (R-ASP)~\cite{Jour:2013_MwMag_Caloz}, inspired from optical analog signal processing (O-ASP)~\cite{BK:2007_Saleh,JOUR:2010_Azana}, might be a complementary, if not a substitutive, solution to future millimeter wave and terahertz technology, where conventional digital signal processing is challenging or inapplicable. Reported R-ASP applications include CRLH compressive receiving~\cite{JOUR:2009_TMTT_Abielmona}, real-time spectrum analysis~\cite{JOUR:2003_TMTT_Laso}, real-time spectrogram analysis~\cite{JOUR:2009_TMTT_Gupta}, temporal expansion for enhanced sampling~\cite{JOUR:2007_TMTT_Schwartz,JOUR:2012_TMTT_Xiang}, temporal compression and reversal~\cite{CONF:2008_IRWS_Schwartz}, real-time spectrum sniffing~\cite{JOUR:2012_MWCL_Nikfal}, SNR enhanced impulse radio transceiving~\cite{JOUR:2014_MWCL_Nikfal}, dispersion code multiple access (DCMA)~\cite{CONF:2015_APS_Gupta}, dispersion-based radio-frequency identification~\cite{CONF:2013_RFID_Perret, JOUR:2011_AWPL_Gupta}, scanning rate uniformization in antenna arrays~\cite{CONF:2015_APS_Zhang}, to name a few.

The core of a R-ASP system is a phaser, a device that provides prescribed group delay frequency responses, $\tau(\omega)$, varying from application to application~\cite{Jour:2013_MwMag_Caloz,JOUR:2015_TMTT_Gupta}. For instance, linear group delay is used in real-time Fourier transformation~\cite{JOUR:2003_TMTT_Laso, JOUR:1999_OPTL_Muriel}, stepped group delay is ideal for distortion-less frequency discrimination~\cite{JOUR:2012_MWCL_Nikfal}, Chebyshev group delay has been used to dispersion-encode radio channels in multiple access communications~\cite{CONF:2015_APS_Gupta}, and a special convex group delay has been synthesized for uniform scanning in an antenna array~\cite{CONF:2015_APS_Zhang}. Moreover, the group delay swing, $\Delta\tau=\tau_\text{max}-\tau_\text{min}$, is a crucial parameter in R-ASP. In applications such as real-time Fourier transforming~\cite{JOUR:2003_TMTT_Laso, JOUR:1999_OPTL_Muriel} and spectrum sniffing~\cite{JOUR:2012_MWCL_Nikfal}, larger $\Delta\tau$ provide better frequency discrimination in time, while in DCMA~\cite{CONF:2015_APS_Gupta}, increasing $\Delta\tau$ results in reduced multi-channel interference, and hence lower bit error rate.

Applications requiring real-time adaptation to dynamic environments would typically require group delay reconfigurability, i.e real-time tunable group delay response ($\tau(\omega)$) or group delay swing ($\Delta\tau$) in R-ASP. One example of such applications is DCMA~\cite{CONF:2015_APS_Gupta}, where the communication channels are essentially time-variant, and hence dynamic communication switching between different pairs of users require real-time reconfiguration of the dispersion codes assigned to each user. In~\cite{JOUR:2004_JLT_Schwelb}, it is shown that, in a single fiber ring resonator~\cite{JOUR:1985_TMTT_Jackson}, varying the loss of the ring results in group delay tuning, where higher loss results in higher group delay swing. However, loss also introduces a notch in the transmission amplitude at the resonance frequency. Moreover, the optical ring resonator in~\cite{JOUR:2004_JLT_Schwelb} requires a forward-wave (co-directional) coupler, which would be excessively long, and hence not practical, at microwave frequencies~\cite{BK:2007_Mongia}. A microwave reconfigurable phaser is reported in~\cite{JOUR:2013_TMTT_Xiang} using a
distributed amplifier to mimic an EBG structure~\cite{Jour:2015_MwMag_Arnedo, JOUR:2003_TMTT_Laso}, but the obtained amplitude response is strongly frequency dependent, which is generally undesirable in R-ASP since this induces distortion~~\cite{Jour:2013_MwMag_Caloz}. In~\cite{JOUR:2011_TMTT_Nikfal}, an amplification feedback loop is added to a C-section phaser to increase the group delay swing. However, the group delay enhancement there is due to the equivalent cascade of $N$ identical phasers, where $N$ is the iteration counts controlled by an SPDT RF switch, which requires external synchronization and design complication. In that system, the amplification in the feedback loop does not contribute to group delay enhancement, but only loss compensation.

We have shown that introducing balanced loss and gain in a conventional C-section phaser~\cite{JOUR:1966_TMTT_Cristal,JOUR:1969_TMTT_Cristal, JOUR:2010_TMTT_Gupta} equally enhances the overall group delay of the phaser while reversing its magnitude response~\cite{CONF:2015_APMC_ZOU}. Furthermore, using equalized loss-gain pair of C-sections results in a perfect phaser, exhibiting flat (all-pass) magnitude and controllable group delay, a response that is impossible in purely passive phasers, where higher group delays are always accompanied with higher loss due to longer time wave trapping, and hence increased dissipation, within the structure. This paper will analytically and experimentally demonstrates the proposed loss-gain pair equalized phasers.

The paper is organized as follows. Section~\ref{SEC:LCResonator} analyzes separate loss or gain C-sections, first with distributed and then with lumped loss or gain. Section~\ref{SEC:CombinedCell} subsequently presents a combined loss-gain pair phaser. Then Sec.~\ref{SEC:DESIGN} and Sec.~\ref{SEC:SYN} describe the design of an equalized loss-gain pair and the experimental demonstration of the resulting all-pass reconfigurable phaser. Finally, a short conclusion is given in Sec.~\ref{SEC:CONCLU}.

\section{Separate loss or gain C-sections}\label{SEC:LCResonator}

\subsection{Distributed Loss or Gain C-Sections}

\begin{figure}[h!t]
   \centering
   \psfragfig*[width=0.3\linewidth, trim={0in 0in 0in 0in}]{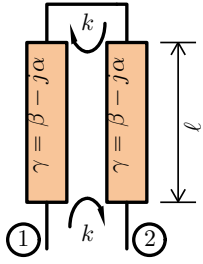}{
    \psfrag{g}[l][l][0.9]{$\gamma=\beta-j\alpha$}
    \psfrag{l}[c][c][0.9]{$\ell$}
    \psfrag{k}[c][c][0.9]{$k$}
    \psfrag{1}[c][c][0.9]{$1$}
    \psfrag{2}[c][c][0.9]{$2$}
  }
   \caption{C-section with physical length $\ell$, maximum coupling coefficient $k$, and isolated transmission line propagation constant $\gamma=\beta-j\alpha$.}
   \label{FIG:ConvCsec}
\end{figure}

A C-section phaser is formed by shorting the end of a backward-wave (contra-directional) coupler, with isolated transmission line propagation constant \mbox{$\gamma=\beta-j\alpha$}~\cite{Jour:2013_MwMag_Caloz}. C-section phasers reported to date are composed of purely \emph{passive} transmission lines, and therefore \mbox{$\alpha>0$}\footnote{The time dependence~$e^{+j\omega t}$ is assumed throughout the paper.}. We will consider here also \emph{active} transmission line C-section phasers, for which \mbox{$\alpha<0$}. Such structures might be engineerable using traveling-wave tube structures or active artificial transmission lines. Figure~\ref{FIG:ConvCsec} shows the general concept of a gain or loss C-section phaser.

\begin{figure}[h!t]
   \centering
   \psfragfig*[width=1\linewidth, trim={0in 0in 0in 0in}]{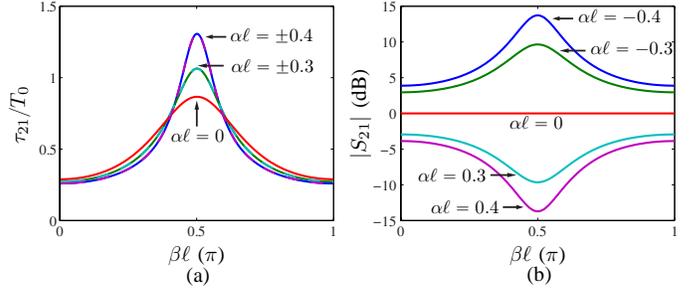}{
    \psfrag{t}[c][c][0.8]{$\beta\ell$ ($\pi$)}
    \psfrag{g}[c][c][0.8]{$\tau_{21}/T_0$}
    \psfrag{a}[c][c][0.8]{$|S_{21}|$ (dB)}
    \psfrag{A}[l][l][0.7]{$\alpha{\ell} = -0.4$}
    \psfrag{B}[l][l][0.7]{$\alpha{\ell} = -0.3$}
    \psfrag{C}[r][r][0.7]{$\alpha{\ell} = 0.3$}
    \psfrag{D}[r][r][0.7]{$\alpha{\ell} = 0.4$}
    \psfrag{E}[c][c][0.7]{$\alpha{\ell} = 0$}
    \psfrag{F}[l][l][0.7]{$\alpha{\ell} =\pm0.4$}
    \psfrag{G}[l][l][0.7]{$\alpha{\ell} =\pm0.3$}
    \psfrag{H}[c][c][0.7]{$\alpha{\ell} =0$}
    \psfrag{c}[c][c][0.8]{(a)}
    \psfrag{d}[c][c][0.8]{(b)}
  }
   \caption{(a)~Normalized group delays, with respect to the period, $T_0$, of the quarter wavelength frequency and (b)~amplitudes, with varying $\alpha\ell$ of a distributed loss or gain C-section, where the coupling factor $k=0.5$.}
   \label{FIG:DistGLResps}
\end{figure}

The transfer function of such a C-section structure takes the general form~\cite{JOUR:2010_TMTT_Gupta}
\begin{equation}\label{EQ:TransFunc_CSec}
\begin{split}
  S_{21} =& \dfrac{1-j\kappa\tan\gamma\ell}{1+j\kappa\tan\gamma\ell}\\
   =&\dfrac{1-\kappa\tanh\alpha\ell-j(\kappa-\tanh\alpha\ell)\tan\beta\ell}
  {1+\kappa\tanh\alpha\ell+j(\kappa+\tanh\alpha\ell)\tan\beta\ell},
\end{split}
\end{equation}
where $\kappa=\sqrt{(1-k)/(1+k)}$, corresponding to the amplitude and the phase
\begin{subequations}
\begin{equation}\label{EQ:S21_AMP_DISTC}
 |S_{21}| = \sqrt{\dfrac{(1-\kappa\tanh\alpha\ell)^2+(\kappa-\tanh\alpha\ell)^2\tan^2\beta\ell}{(1+\kappa\tanh\alpha\ell)^2+(\kappa+\tanh\alpha\ell)^2\tan^2\beta\ell}},
\end{equation}
\begin{equation}\label{EQ:S21_PHS_DISTC}
\begin{split}
  \angle S_{21} = & -\arctan{\left(\dfrac{\kappa-\tanh\alpha\ell}{1-\kappa\tanh\alpha\ell}\tan\beta\ell\right)}\\
  & -\arctan{\left(\dfrac{\kappa+\tanh\alpha\ell}{1+\kappa\tanh\alpha\ell}\tan\beta\ell\right)},
  \end{split}
\end{equation}
 \end{subequations}

\noindent respectively. Inspecting~\eqref{EQ:S21_AMP_DISTC} and~\eqref{EQ:S21_PHS_DISTC} reveals that reversing the sign of $\alpha$ reverses $|S_{21}|$ and keeps $\phi_{21}$ unchanged, i.e.
\begin{subequations}
  \begin{equation}\label{EQ:DIST_AMP}
    |S_{21}(-\alpha)|=\dfrac{1}{|S_{21}(\alpha)|},
  \end{equation}
  \begin{equation}\label{EQ:DIST_PHS}
    \angle S_{21}(-\alpha) = \angle S_{21}(\alpha),
  \end{equation}
\end{subequations}

\noindent respectively. Equations~\eqref{EQ:DIST_AMP} and~\eqref{EQ:DIST_PHS} state that equalized distributed loss ($|\alpha|$) and gain ($-|\alpha|$) yields $0$~dB-symmetric amplitudes, $20\log(|S_{21}(-\alpha)|)=-20\log(|S_{21}(\alpha)|)$, and identical group delays, $\tau_{21}(-\alpha)=\tau_{21}(\alpha)$, from $\tau_{21}=-\partial\angle S_{21}/\partial\omega$, as plotted in~\figref{FIG:DistGLResps}. The tuning effect of distributed loss and gain on group delay will be later shown to allow real-time dispersion engineering. However, \emph{distributed} loss and gain profiles would be difficult to engineer, and we therefore now turn to \emph{lumped} loss and gain inclusions, where the same conclusions will be shown to hold.

\subsection{Lumped Loss or Gain C-Sections}\label{SEC:PRINP:PROVE}

\begin{figure}[h!t]
   \centering
   \psfragfig*[width=0.55\linewidth, trim={0in 0in 0in 0in}]{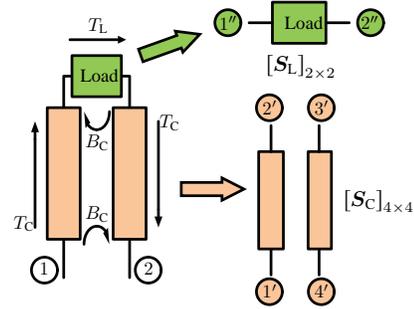}{
    \psfrag{L}[c][c][0.7]{Load}
    \psfrag{l}[c][c][0.7]{Load}
    \psfrag{A}[c][c][0.8]{$\left[\bm{S}_\text{L}\right]_{2\times2}$}
    \psfrag{B}[l][l][0.8]{$\left[\bm{S}_\text{C}\right]_{4\times4}$}
    \psfrag{5}[c][c][0.7]{$1$}
    \psfrag{6}[c][c][0.7]{$2$}
    \psfrag{7}[c][c][0.7]{$1''$}
    \psfrag{8}[c][c][0.7]{$2''$}
    \psfrag{1}[c][c][0.7]{$1'$}
    \psfrag{2}[c][c][0.7]{$2'$}
    \psfrag{3}[c][c][0.7]{$3'$}
    \psfrag{4}[c][c][0.7]{$4'$}
    \psfrag{C}[c][c][0.7]{$B_\text{C}$}
    \psfrag{T}[c][c][0.7]{$T_\text{C}$}
    \psfrag{R}[c][c][0.7]{$T_\text{L}$}

  }
   \caption{Loaded C-section consisting of a coupled-line coupler, represented by a $4\times4$ scattering matrix, $\left[S_\text{C}\right]_{4\times4}$, and a load represented by a $2\times2$ scattering matrix, $\left[S_\text{L}\right]_{2\times2}$.}
   \label{FIG:LoadedCsec}
\end{figure}

We thus now consider the lumped loss or gain C-section phaser shown in \figref{FIG:LoadedCsec}, where the coupler is assumed, in first approximation, to be lossless and terminated at one end by a load. The coupler lossless approximation is very reasonable since dispersion loss is typically much larger than transmission line conductive or dielectric losses~\cite{Jour:2013_MwMag_Caloz}.
Assuming perfect load matching, the scattering matrix of the load is
\begin{subequations}\label{EQ:SP_Load}
    \begin{equation}\label{EQ:SP_Matrix_Load}
        \left[\bm{S}_\text{load}\right]_{2\times2}=
        \left[
        \begin{array}{cc}
            0       &   T_\text{L}\\
        T_\text{L}  &   0
        \end{array}
        \right],
    \end{equation}
    \text{where}
    \begin{equation}\label{EQ:SP_Trans_Load}
        T_\text{L}(\omega)=A_\text{L}(\omega)e^{j\phi_\text{L}(\omega)},
    \end{equation}
\end{subequations}
is the load transfer function. Further assuming coupler perfect matching and isolation, the scattering matrix of the coupler is
\begin{subequations}\label{EQ:SP_Cpl}
    \begin{equation}\label{EQ:SP_Matrix_Cpl}
        \left[\bm{S}_\text{C}\right]_{4\times4}=
        \left[
        \begin{array}{cccc}
            0       &   T_\text{C}  &       0       &   B_\text{C}  \\
        T_\text{C}  &       0       &   B_\text{C}  &       0       \\
            0       &   B_\text{C}  &       0       &   T_\text{C}  \\
        B_\text{C}  &       0       &   T_\text{C}  &       0
        \end{array}
        \right],
    \end{equation}
    \text{where}
    \begin{equation}\label{EQ:SP_Bwc_Cpl}
        B_\text{C}(\omega)=\dfrac{jk\sin\theta}{\sqrt{1-k^2}\cos\theta+j\sin\theta},
    \end{equation}
    \begin{equation}\label{EQ:SP_Thr_Cpl}
        T_\text{C}(\omega)=\dfrac{\sqrt{1-k^2}}{\sqrt{1-k^2}\cos\theta+j\sin\theta},
    \end{equation}
\end{subequations}
are the backward coupling and through transfer functions, respectively~\cite{BK:2007_Mongia}, with parameters $k$ and $\theta = \pi\omega/2\omega_0$ being the maximum coupling, occurring at the quarter-wavelength frequency $\omega_0$, and the electrical length of the coupler at that frequency, respectively.

The backward coupling introduces a feedback loop from the load output port $2''$ to the load input port $1''$ (see Fig.~\ref{FIG:LoadedCsec}), leading to multiple scattering at port $2$ of the loaded C-section. Assuming unity excitation to the loaded C-section port $1$ and accounting for the multiple scattering events, one finds
\begin{equation}\label{EQ:S21_IdealLoadedCSec}
\begin{split}
  S_{21}=&S_{12}\\
  =&B_\text{C}+T_cT_\text{L}T_\text{C} + T_\text{C}T_\text{L}B_\text{C}T_\text{L}T_\text{C}\\
   &+ T_\text{C}T_\text{L} B_\text{C}T_\text{L}B_\text{C}T_\text{L}T_\text{C} + \ldots\\
   =& B_\text{C}+T_\text{C}^2T_\text{L}\sum_{n=0}^{\infty}(B_\text{C}T_\text{L})^n.
\end{split}
\end{equation}
If $|B_\text{C}T_\text{L}|\ge1$, the geometric series~\eqref{EQ:S21_IdealLoadedCSec} diverges, corresponding to an oscillatory (unstable) regime. Therefore, the condition $|B_\text{C}T_\text{L}|<1$ must be enforced for stability, in which case~\eqref{EQ:S21_IdealLoadedCSec} reduces to
\begin{equation}\label{EQ:S21_IdealLoadedCSec_converged}
  S_{21}=S_{12}=B_\text{C}+\dfrac{T_\text{C}^2T_\text{L}}{1-B_\text{C}T_\text{L}},
\end{equation}
with
\begin{equation}\label{EQ:LoopGainCond1_AMP}
\left|T_\text{L}(\omega)\right|=A_\text{L}(\omega)<\dfrac{1}{\left|B_\text{C}(\omega)\right|}.
\end{equation}
Moreover, according to~\cite{JOUR:2013_APMag_Zhang}, the maximum of the group delay occurs at the resonance frequency, $\omega_\text{p}$, where the multiple scattered waves add in phase. This is equivalent to the following phase condition
\begin{equation}\label{EQ:LoopGainCond1_PHA}
  \angle{\left[T_\text{L}(\omega_\text{p})B_\text{C}(\omega_\text{p})\right]}=\phi_\text{L}(\omega_\text{p})+\angle{B_\text{C}(\omega_\text{p})}=2n\pi,
\end{equation}
where $n$ is an integer. Equation~\eqref{EQ:LoopGainCond1_PHA} suggests that, given a coupler with known $\angle{B_c(\omega)}$, the resonance frequency, $\omega_\text{p}$, may be tuned by varying the load transmission phase $\phi_\text{L}$.

\begin{figure}[h!t]
   \centering
   \psfragfig*[width=1\linewidth]{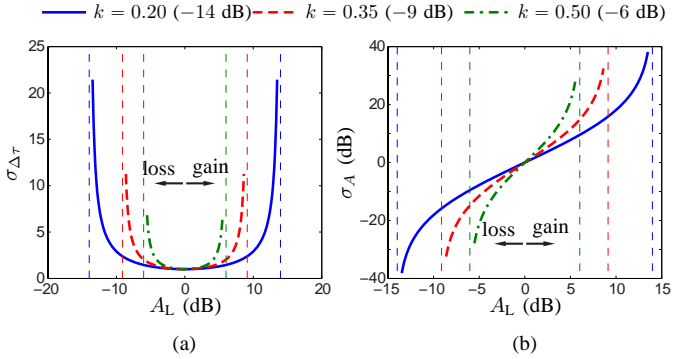}{
    \psfrag{A}[c][c][0.8]{$A_\text{L}$ (dB)}
    \psfrag{H}[c][c][0.8]{$\sigma_{\Delta\tau}$}
    \psfrag{S}[c][c][0.8]{$\sigma_{A}$ (dB)}
    \psfrag{X}[l][l][0.75]{$k=0.20$ ($-14$ dB)}
    \psfrag{Y}[l][l][0.75]{$k=0.35$ ($-9$ dB)}
    \psfrag{Z}[l][l][0.75]{$k=0.50$ ($-6$ dB)}
    \psfrag{G}[c][c][0.8]{gain}
    \psfrag{L}[c][c][0.8]{loss}
    \psfrag{C}[tc][mc][0.8]{(a)}
    \psfrag{D}[tc][mc][0.8]{(b)}
  }
   \caption{Tuning effect of the load transmission amplitude, $A_\text{L}$ in dB, on (a)~the loaded C-section group delay swing and (b)~the transmission amplitude, respectively, with tuning range of $A_\text{L}$ defined by $k$ as the lower limit (left dashed line) and $1/k$ as the upper limit (right dashed line). Range: $A_\text{L}\in{[k+0.5\text{ dB}, 1/k-0.5\text{ dB}]}$.}
   \label{FIG:TunningEffect_AMP}
\end{figure}

Inserting~\eqref{EQ:SP_Trans_Load},~\eqref{EQ:SP_Bwc_Cpl} and~\eqref{EQ:SP_Thr_Cpl} into~\eqref{EQ:S21_IdealLoadedCSec} yields
\begin{subequations}
  \begin{equation}\label{EQ:S21_IdealLoadedCSec_Expl}
  \begin{split}
    S_{21}=&T_\text{L}\dfrac{\sqrt{1-k^2}\cot\theta+{k}/{A_\text{L}}\sin\phi_\text{L}-j\left(1-{k}/{A_\text{L}}\cos\phi_\text{L}\right)}
    {\sqrt{1-k^2}\cot\theta+{kA_\text{L}}\sin\phi_\text{L}+j\left(1-kA_\text{L}\cos\phi_\text{L}\right)}\\
    =& \left|S_{21}\right|e^{j\phi_{21}},
  \end{split}
  \end{equation}
  \text{where}
\begin{equation}\label{EQ:S21_Mag_IdealLoadedCSec_Expl}
\begin{split}
  &\left|S_{21}\right|=\\
  &A_\text{L}\sqrt{\dfrac{\left(\sqrt{1-k^2}\cot\theta+{k}/{A_\text{L}}\sin\phi_\text{L}\right)^2+\left(1-{k}/{A_\text{L}}\cos\phi_\text{L}\right)^2}
  {\left(\sqrt{1-k^2}\cot\theta+kA_\text{L}\sin\phi_\text{L}\right)^2+\left(1-kA_\text{L}\cos\phi_\text{L}\right)^2}},
\end{split}
\end{equation}
\begin{equation}\label{EQ:S21_Pha_IdealLoadedCSec_Expl}
\begin{split}
  \angle S_{21}=&-\arctan\dfrac{1-{k}/{A_\text{L}}\cos\phi_\text{L}}{\sqrt{1-k^2}\cot\theta+{k}/{A_\text{L}}\sin\phi_\text{L}}\\
            &-\arctan\dfrac{1-kA_\text{L}\cos\phi_\text{L}}{\sqrt{1-k^2}\cot\theta+kA_\text{L}\sin\phi_\text{L}}\\
            &+\phi_\text{L},
\end{split}
\end{equation}
\end{subequations}
Inspecting~\eqref{EQ:S21_Mag_IdealLoadedCSec_Expl} and~\eqref{EQ:S21_Pha_IdealLoadedCSec_Expl} reveals that reversing the load transmission amplitude, i.e. $A_\text{L}\rightarrow1/A_\text{L}$, reverses the loaded C-section transfer function amplitude, $|S_{21}|$, while keeping the transmission phase $\angle S_{21}$, and hence also the group delay, $\tau_{21}$, unchanged. Therefore, the lumped loss or gain loaded C-section indeed exhibits the same fundamental properties as its distributed counterpart.
For simplicity, we next assume that the load transmission is of constant amplitude and linear phase, or is non-dispersive, i.e. \mbox{$A_\text{L}(\omega) = G (G>1)$} or  \mbox{$A_\text{L}(\omega)=L (L<1)$}, so that
\begin{equation}\label{EQ:LoadingNetworkPhase}
\phi_\text{L} = -\tau_\text{L}\omega,
\end{equation}
where $\tau_\text{L}$ is the load transmission delay. we may now parametrically study the behavior of the loaded C-section. Assuming the first resonance frequency, $\omega_\text{p}=\omega_0$, the maximum of the loaded C-section group delay corresponds then to
\begin{equation}\label{EQ:S21_GD_IdealLoadedCSec_CF_Complete}
\begin{split}
\tau_{21}^\text{max}(A_\text{L},\omega)=& \tau_{21}(A_\text{L}, \omega_0) =-\left.\dfrac{\partial{\phi_{21}}}{\partial{\omega}}\right|_{\omega=\omega_0}\\
=&\dfrac{T_0\sqrt{1-k^2}}{4}\left(\dfrac{1}{1-{k}/{A_\text{L}}}+\dfrac{1}{1-kA_\text{L}}\right)\\
&+\tau_\text{L}{k}\left(\dfrac{{1}/{A_\text{L}}}{1-{k}/{A_\text{L}}}+\dfrac{A_\text{L}}{1-kA_\text{L}}\right)\\
&+\tau_\text{L},
\end{split}
\end{equation}
where \mbox{$\phi_\text{L}(\omega_0)=2n\pi-\angle{B_\text{C}(\omega_0)}=2n\pi$}, with $\angle{B_\text{C}(\omega_0)}=0$~\cite{BK:2007_Mongia} has been used. Further assuming $\tau_\text{L}=0$ reduces~\eqref{EQ:S21_GD_IdealLoadedCSec_CF_Complete} to
\begin{equation}\label{EQ:S21_GDMax}
  \tau_{21}^\text{max}(A_\text{L},\omega) = \dfrac{T_0\sqrt{1-k^2}}{4}\left(\dfrac{1}{1-{k}/{A_\text{L}}}+\dfrac{1}{1-kA_\text{L}}\right),
\end{equation}
with $T_0 = 1/f_0$ being the period at the quarter-wavelength frequency.
The minimum loaded C-section group delay occurring at the first anti-resonance frequency, $2\omega_\text{p} = 2\omega_0$, is
\begin{equation}\label{EQ:S21_GDMin}
\begin{split}
  \tau_{21}^\text{min}(A_\text{L},\omega)=& \tau_{21}(A_\text{L}, 2\omega_0) =-\left.\dfrac{\partial{\phi_{21}}}{\partial{\omega}}\right|_{\omega=2\omega_0, \tau_\text{L} = 0}\\
  =& \dfrac{T_0}{4\sqrt{1-k^2}}\left(2-kA_\text{L}-\dfrac{k}{A_\text{L}}\right),
\end{split}
\end{equation}
The difference of the last two relations corresponds to the group delay swing
\begin{equation}\label{EQ:GDSwing}
\begin{split}
  \Delta\tau_{21}(A_\text{L}) &= \tau_{21}^\text{max}(A_\text{L}, \omega)-\tau_{21}^\text{min}(A_\text{L}, \omega)\\
  & = \dfrac{T_0k\left(A_\text{L}+1/A_\text{L}-2k\right)}{4\sqrt{1-k^2}}\left(\dfrac{1}{1-k/A_\text{L}}+\dfrac{1}{1-kA_\text{L}}\right).
\end{split}
\end{equation}
We may at this point define the loaded C-section transmission group delay swing and amplitude tuning factors
\begin{subequations}
  \begin{equation}\label{EQ:S21_GDEnh_IdealLOadedCSec_CF}
\begin{split}
  \sigma_{\Delta\tau}(A_\text{L}) &=  \dfrac{\Delta\tau_{21}(A_\text{L})}{\Delta\tau_{21}(A_\text{L}=1)}\\
  &=\dfrac{A_\text{L}+1/A_\text{L}-2k}{4}\left(\dfrac{1}{1-{k}/{A_\text{L}}}+\dfrac{1}{1-kA_\text{L}}\right)
\end{split}
\end{equation}
\text{and}
\begin{equation}\label{EQ:S21_MAG_IdealLoadedCSec_CF}
\sigma_{A}(A_\text{L}) = \dfrac{|S_{21}(A_\text{L}, \omega_0)|_\text{}}{|S_{21}(A_\text{L}=1, \omega_0)|} = \left|S_{21}(A_\text{L}, \omega_0)\right|=\left|\dfrac{A_\text{L}-k}{1-kA_\text{L}}\right|.
\end{equation}
\end{subequations}
Figure~\ref{FIG:TunningEffect_AMP} shows $\sigma_{\Delta\tau}(A_\text{L})$ and $\sigma_{A}(A_\text{L})$ with three different coupling factors $k$. We see that $\sigma_{\Delta\tau}(A_\text{L})$ and $\sigma_{A}(A_\text{L})$ (in~dB) are even and odd function of $A_\text{L}$ (in~dB), respectively, which means that a balanced pair of load loss and load gain have same tuning effect on group delay swing, while opposite tuning effect on amplitude. Moreover, smaller $k$ gives higher group delay swing tuning range but at the cost of using higher loss or gain load and hence consuming more power. Also note that $A_\text{L}$ going above upper limit $1/k$ leads to oscillation, which should be avoided, while going below lower limit $k$ results in negative group delay, which has been presented in~\cite{JOUR:2013_APMag_Zhang}.

\begin{figure}[h!t]
   \centering
   \psfragfig*[width=1\linewidth]{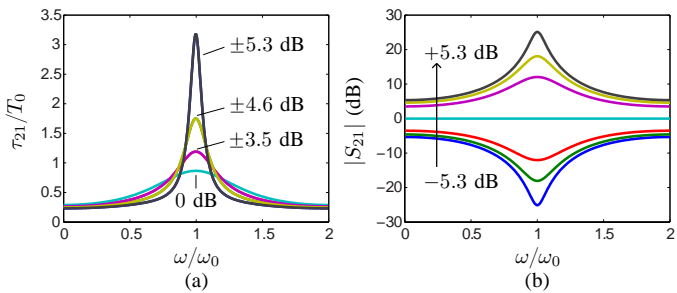}{
    \psfrag{A}[c][c][0.8]{$|S_{21}|$ (dB)}
    \psfrag{T}[c][c][0.8]{$\tau_{21}/T_0$}
    \psfrag{F}[c][c][0.8]{$\omega/\omega_0$}
    \psfrag{x}[Tc][bc][0.8]{(a)}
    \psfrag{z}[Tc][bc][0.8]{(b)}
    \psfrag{a}[c][c][0.8]{$0$ dB}
    \psfrag{b}[l][l][0.8]{$\pm3.5$ dB}
    \psfrag{c}[l][l][0.8]{$\pm4.6$ dB}
    \psfrag{d}[l][l][0.8]{$\pm5.3$ dB}
    \psfrag{e}[l][l][0.8]{$-5.3$ dB}
    \psfrag{f}[l][l][0.8]{$+5.3$ dB}
  }
  \caption{Loaded C-section (a)~normalized group delays $\tau_{21}/T_0$ and (b)~transmission amplitudes $|S_{21}|$, with maximum coupling factor $k=-6$ dB ($k=0.5$) at $\omega_0$, and varying load transmission amplitudes  $A_\text{L}=\{0, \pm3.5, \pm4.6, \pm5.3\}$ dB. }
   \label{FIG:Resp_IdealLoadedCSection}
\end{figure}

Figure~\ref{FIG:Resp_IdealLoadedCSection} shows the loaded C-section transmission group delays and amplitudes for coupling coefficient $k=0.5$. We see that the equalized loss and gain pair exhibits identical group delay and symmetric amplitudes about $|S_{21}|=0$~dB.

\begin{figure}[h!t]
   \centering
   \psfragfig*[width=1\linewidth]{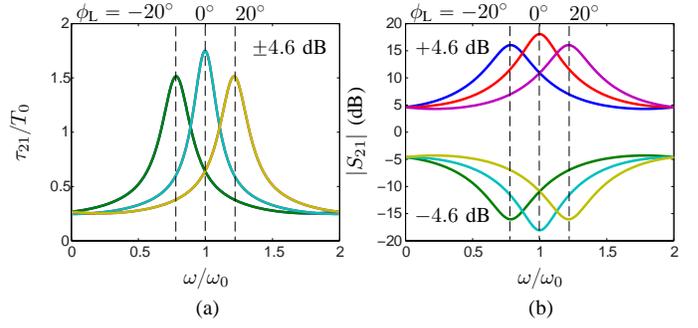}{
    \psfrag{A}[c][c][0.8]{$|S_{21}|$ (dB)}
    \psfrag{T}[c][c][0.8]{$\tau_{21}/T_0$}
    \psfrag{F}[c][c][0.8]{$\omega/\omega_0$}
    \psfrag{G}[c][c][0.8]{(a)}
    \psfrag{B}[c][c][0.8]{(b)}
    \psfrag{C}[l][l][0.8]{$\pm4.6$ dB}
    \psfrag{E}[l][l][0.8]{$+4.6$ dB}
    \psfrag{D}[l][l][0.8]{$-4.6$ dB}
    \psfrag{X}[r][r][0.8]{$\phi_\text{L}=-20^\circ$}
    \psfrag{Y}[c][c][0.8]{$0^\circ$}
    \psfrag{Z}[l][l][0.8]{$20^\circ$}
  }
   \caption{Loaded C-section (a) normalized group delays $\tau_{21}/T_0$ and (b) transmission amplitudes $|S_{21}|$, with maximum coupling factor $k=6$ dB ($k=0.5$) at $\omega_0$ and $A_\text{L} = \pm4.6$ dB, and varying load transmission phase $\phi_\text{L}=\{-20^\circ, 0^\circ, 20^\circ\}$.  }
   \label{FIG:Resp_IdealLoadedCSection2}
\end{figure}

Apart from tuning the height of group delay peak, the position of group delay peak, $\omega_\text{p}$, or resonance frequency, can be also tuned by varying the load transmission phase, $\phi_\text{L}$, as shown in~\figref{FIG:Resp_IdealLoadedCSection2}. The value of $\omega_\text{p}$ is determined by using~\eqref{EQ:LoopGainCond1_PHA}.

\section{Combined Loss-Gain Equalized Pair}\label{SEC:CombinedCell}

\begin{figure}[h!t]
   \centering
   \psfragfig*[width=0.4\linewidth, trim={0in 0in 0in 0in}]{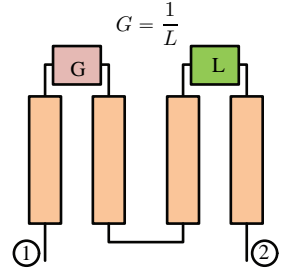}{
    \psfrag{g}[c][c][0.8]{G}
    \psfrag{l}[c][c][0.8]{L}
    \psfrag{e}[c][c][0.8]{$G=\dfrac{1}{L}$}
    \psfrag{1}[c][c][0.8]{1}
    \psfrag{2}[c][c][0.8]{2}
  }
   \caption{Proposed all-pass loss-gain equalized pair formed by  serially
connecting a loss loaded C-section and a gain loaded C-section, where the gain is the reverse of the loss, $G=1/L$. }
   \label{FIG:EqualizedPhaser}
\end{figure}

\begin{figure}[h!t]
   \centering
   \psfragfig*[width=1\linewidth, trim={0in 0in 0in 0in}]{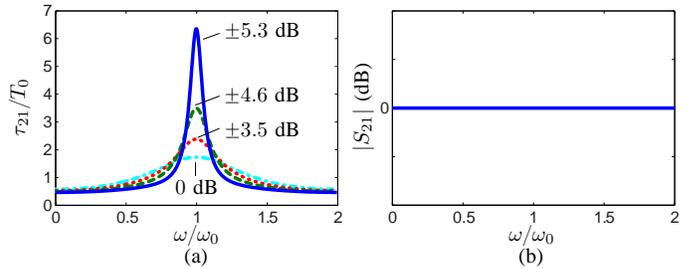}{
    \psfrag{A}[c][c][0.8]{$|S_{21}|$ (dB)}
    \psfrag{T}[c][c][0.8]{$\tau_{21}/T_0$}
    \psfrag{F}[c][c][0.8]{$\omega/\omega_0$}
    \psfrag{a}[l][l][0.8]{$\pm5.3$ dB}
    \psfrag{b}[l][l][0.8]{$\pm4.6$ dB}
    \psfrag{c}[l][l][0.8]{$\pm3.5$ dB}
    \psfrag{d}[c][c][0.8]{$0$ dB}
    \psfrag{x}[c][c][0.8]{(a)}
    \psfrag{z}[c][c][0.8]{(b)}
  }
   \caption{Loss-gain equalized pair (a) normalized group delay $\tau_{21}/T_0$ and (b) all-pass transmission amplitude $|S_{21}|$, with maximum coupling factor $k=0.5$ at $\omega_0$ and different load transmission amplitudes $A_\text{L} =\{0, \pm3.5, \pm4.6, \pm5.3\}$ dB. }
   \label{FIG:Resp_Equalized}
\end{figure}

An all-pass loss-gain equalized pair is formed by serially connecting a loss C-section and a gain C-section, as shown in~\figref{FIG:EqualizedPhaser}, with appropriately tune gain, $G$, and loss, $L$, such that $G=1/L$. The group delay of the resulting loss-gain pair is twice that of a single loss or gain loaded C-section phaser, as shown in~\figref{FIG:Resp_Equalized}(a), while the transmission amplitude becomes all-pass [\figref{FIG:Resp_Equalized}(b)].

The analysis performed so far assumes an ideal (lossless, perfectly matched and perfectly isolated) system. As a result, the loaded C-section is stable as long as condition~\eqref{EQ:LoopGainCond1_AMP} is satisfied. In reality, the loaded C-section phaser may still become unstable due to non-ideal factors, such as coupler forward-wave coupling (imperfect isolation) and load mismatch, since such non-idealities create wave paths that have not been accounted for in~\eqref{EQ:LoopGainCond1_AMP}. A detailed stability analysis of a real C-section phaser is beyond the scope of this paper, and will be presented elsewhere. However, maximizing isolation and matching clearly appears to represent important design considerations from the viewpoint of stability.

\section{Design of Loss-Gain Pair}\label{SEC:DESIGN}
\subsection{Microstrip Coupler}

The C-section phaser is implemented here in microstrip technology for easiest fabrication and testing. Unfortunately, due to their imperfect transverse electromagnetic nature, and corresponding unequal even and odd mode phase velocities ($v_\text{o}>v_\text{e}$), microstrip couplers suffer from relatively poor isolation~\cite{BK:2007_Mongia}. To minimize the aforementioned subsequent risks of instability, one should thus increase the natural isolation of the coupler. Corresponding equalization of even and odd velocities may be achieved by different approaches, such as using wiggly transmission lines~\cite{JOUR:1989_TMTT_Uysal}, inductive compensation~\cite{JOUR:2010_TMTT_Lee} or capacitive compensation~\cite{JOUR:1999_TMTT_Dydyk}, etc. We use here capacitive compensation, which consists in inserting a coupling enhancing capacitance in the gap between the two transmission lines.

\begin{figure}[h!t]
\centering
   \psfragfig*[width=0.555\linewidth, trim={0in 0in 0in 0in}]{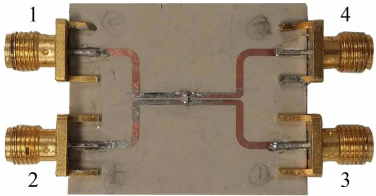}{
   \psfrag{1}[c][c][0.8]{\textcolor{blue}{1}}
   \psfrag{2}[c][c][0.8]{\textcolor{blue}{2}}
   \psfrag{3}[c][c][0.8]{\textcolor{blue}{3}}
   \psfrag{4}[c][c][0.8]{\textcolor{blue}{4}}}
   \caption{Fabricated microstrip coupler (Rogers RO6010, \mbox{$\epsilon_r = 10.2$}, $0.5$ oz cladding, $50$ mil substrate).}\label{FIG:PhotoCpl}
\end{figure}

\begin{figure}[h!t]
   \centering
   \psfragfig*[width=0.65\linewidth, trim={0in 0in 0in 0in}]{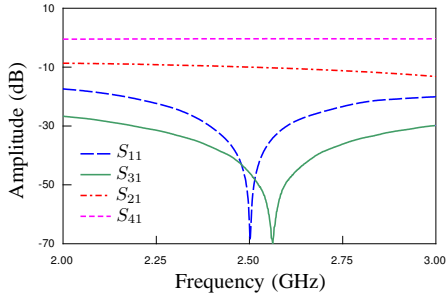}{
   \psfrag{f}[c][c][0.8]{Frequency (GHz)}
   \psfrag{A}[c][c][0.8]{Amplitude (dB)}
   \psfrag{i}[l][l][0.7]{$S_{11}$}
   \psfrag{r}[l][l][0.7]{$S_{31}$}
   \psfrag{c}[l][l][0.7]{$S_{21}$}
   \psfrag{t}[l][l][0.7]{$S_{41}$}
    }
   \caption{Measured S-parameters of the microstrip coupler shown in~\figref{FIG:PhotoCpl}. }
   \label{FIG:Resp_HiIsoCpl}
\end{figure}

The fabricated microstrip coupler is shown in~\figref{FIG:PhotoCpl} while Fig.~\ref{FIG:Resp_HiIsoCpl} shows the corresponding measured response, with best matching and isolation reached near $2.5$ GHz and corresponding $-10$~dB coupling. Therefore we will choose operation frequency around $2.5$~GHz in the design of phaser later.

\subsection{Variable Loss-Gain Load}

\begin{figure}[h!t]
   \centering
   \psfragfig*[width=0.5\linewidth, trim={0in 0in 0in 0in}]{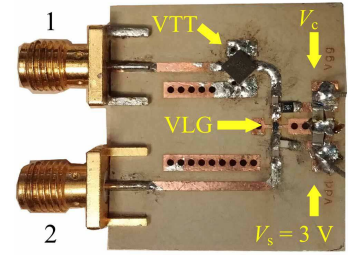}{
   \psfrag{D}[r][r][0.8]{\textcolor{yellow}{$V_\text{s}=3$ V}}
   \psfrag{G}[c][c][0.8]{\textcolor{yellow}{$V_\text{c}$}}
   \psfrag{T}[r][r][0.8]{\textcolor{yellow}{VLG}}
   \psfrag{A}[l][l][0.8]{\textcolor{yellow}{ATT}}
   \psfrag{1}[c][c][0.8]{\textcolor{blue}{1}}
   \psfrag{2}[c][c][0.8]{\textcolor{blue}{2}}}
    \caption{Fabricated variable loss or gain load (same substrate as in \figref{FIG:PhotoCpl}) composed a VLG (Avago VMMK-3503~\cite{DS_VGA}: internal matching, $0.5-18$~GHz operation, 10~dB gain to $13$~dB loss tuning range ($23$~dB interval) by varying $V_\text{c}$ from $1.8$~V to $0$~V) and a fixed stabilization and matching enhancement attenuator (Minicircuit GAT-4+: 4~dB attenuation).}\label{FIG:PhotoLoad}
\end{figure}

The fabricated loss-gain load is shown in~\figref{FIG:PhotoLoad} with parts specifications. It is composed of the series connection of an internally matched variable loss-gain~(VLG) chip, whose loss-gain is tuned by varying the control bias voltage, and a fixed stabilization and matching enhancement attenuator~(ATT). Both the VLG internal matching and the ATT matching and attenuation contribute to the stability of the overall device discussed in Sec.~\ref{SEC:CombinedCell}.

Figure~\ref{FIG:Resp_HiIsoCpl} shows the corresponding measured response. The maximum measured gain is limited by the attenuator to the level of $6.7$~dB, which lies in the stability region prescribed by~\eqref{EQ:LoopGainCond1_AMP}, namely $|T_{\text{L}{F}}(2.5\text{ GHz})|^\text{max}=1/|B_c(2.5\text{ GHz})|=10$~dB.

\begin{figure}[h!t]
   \centering
   \psfragfig*[width=1.0\linewidth, trim={0in 0in 0in 0in}]{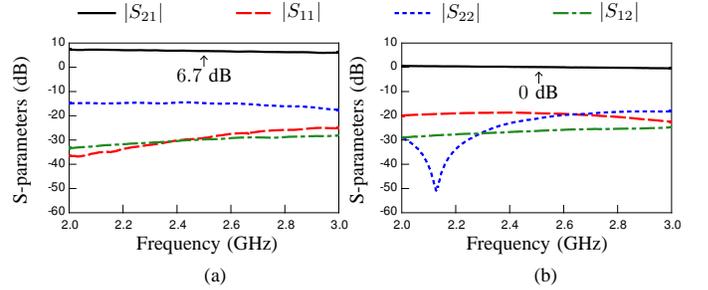}{
   \psfrag{F}[c][c][0.75]{Frequency (GHz)}
   \psfrag{a}[l][l][0.75]{$|S_{21}|$}
   \psfrag{b}[l][l][0.75]{$|S_{11}|$}
   \psfrag{c}[l][l][0.75]{$|S_{22}|$}
   \psfrag{d}[l][l][0.75]{$|S_{12}|$}
   \psfrag{e}[l][l][0.75]{(a)}
   \psfrag{r}[l][l][0.75]{(b)}
   \psfrag{A}[c][c][0.75]{$6.7$ dB}
   \psfrag{B}[c][c][0.75]{$0$ dB}
    \psfrag{S}[c][c][0.75]{S-parameters (dB)}}
   \caption{Measured amplitude responses of the load shown in~\figref{FIG:PhotoLoad} for (a)~$|T_{\text{L}}(2.5 \text{ GHz})|=6.7$~dB and (b)~$|T_{\text{L}}(2.5\text{ GHz}|=0$~dB.}
   \label{FIG:Resp_GainLoad}
\end{figure}

\subsection{Numerical and Experimental Demonstration of Loss-Gain C-Sections}

\begin{figure}[h!t]
 \centering

 \psfrag{a}[l][l][0.8]{$|T_{\text{L}}(2.5\text{ GHz})|=6.7$  dB}
 \psfrag{b}[l][l][0.8]{$|T_{\text{L}}(2.5\text{ GHz})|=0$  dB}
 \psfrag{F}[c][c][0.8]{Frequency (GHz)}
\subfigure[]{
   \psfragfig*[width=1\linewidth, trim={0in 0in 0in 0in}]{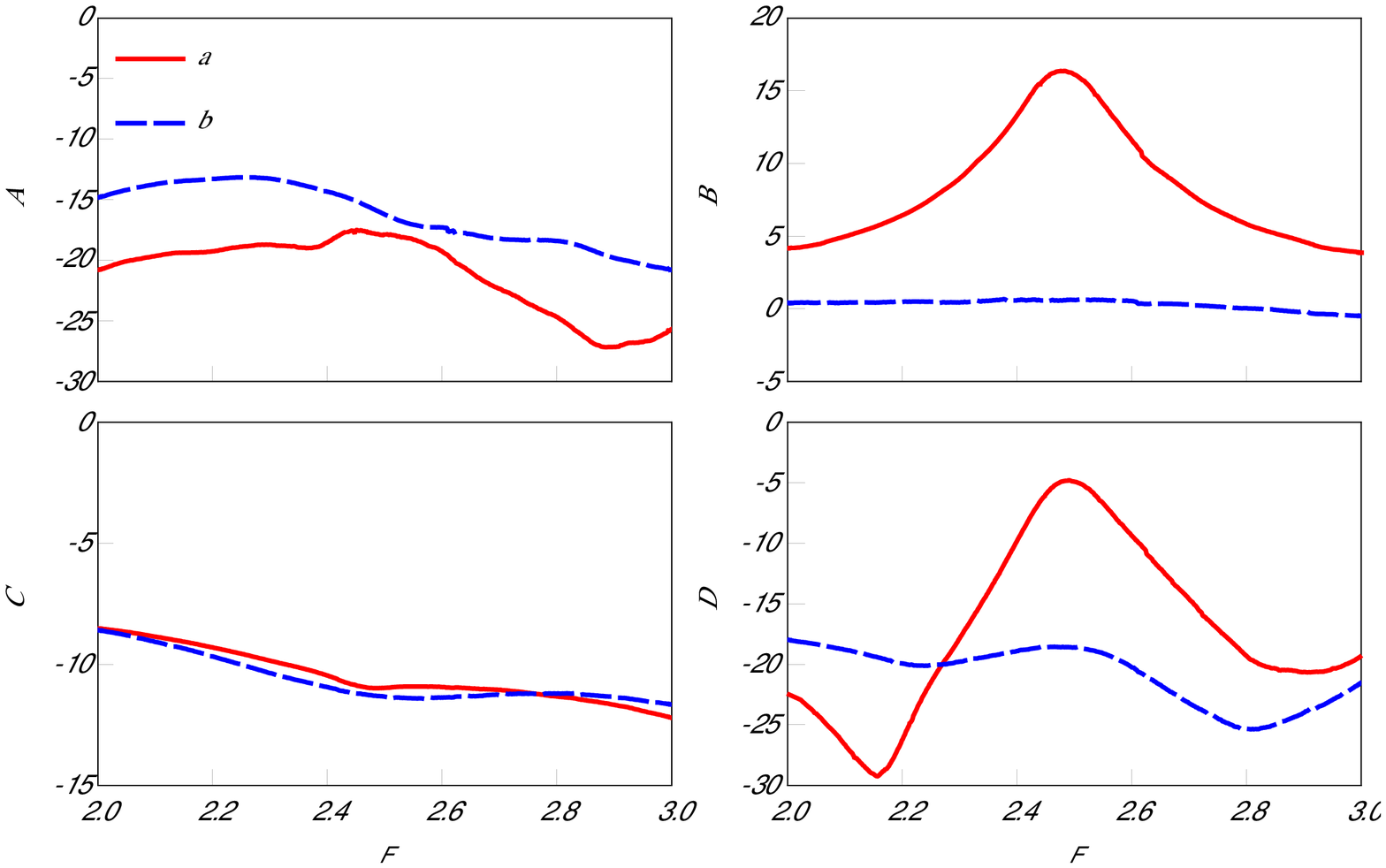}{
    \psfrag{A}[c][c][0.8]{$|S_{11}|$ (dB)}
    \psfrag{B}[c][c][0.8]{$|S_{21}|$ (dB)}
    \psfrag{C}[c][c][0.8]{$|S_{12}|$ (dB)}
    \psfrag{D}[c][c][0.8]{$|S_{22}|$ (dB)}
  }
   \label{FIG:Resp_SimSingleAMP}
   }

   \subfigure[]{
   \psfragfig*[width=0.6\linewidth, trim={0in 0in 0in 0in}]{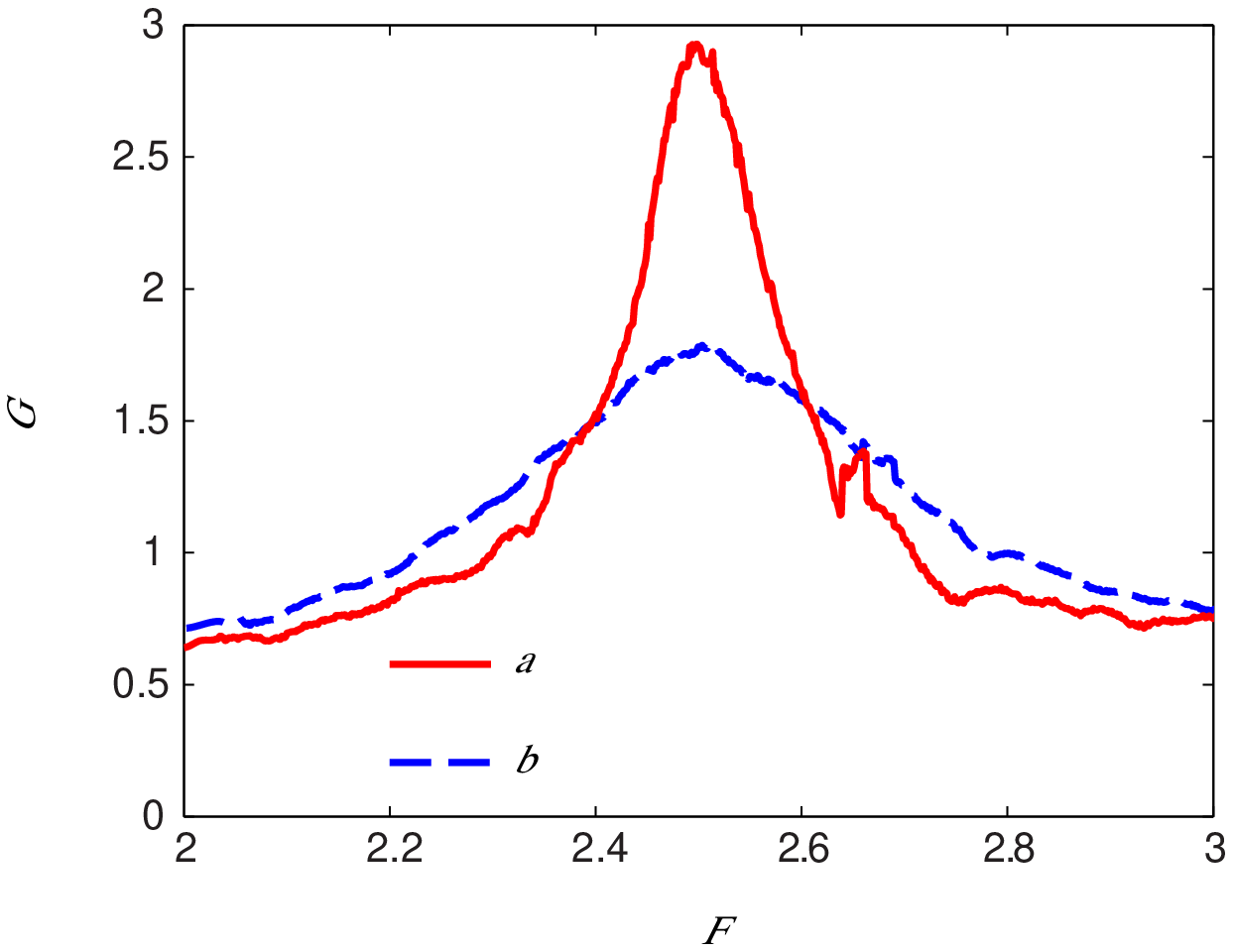}{
    \psfrag{G}[c][c][0.8]{Group delay $\tau_{21}$ (ns)}
  }
   \label{FIG:Resp_SimSingleGD}}
   \caption{Simulated (a)~amplitudes and (b)~group delays of a loaded C-section phasers with imported coupler and load experimental models corresponding to~\figref{FIG:Resp_HiIsoCpl} and~\figref{FIG:Resp_GainLoad}, respectively.}
   \label{FIG:Resp_SimSingle}
\end{figure}

The performance of the loss or gain C-section phaser may be predicted using a commercial RF circuit simulator by importation of the measured coupler and load responses into two-port and four-port scattering models, respectively. Figure~\ref{FIG:Resp_SimSingleAMP} shows the corresponding responses. The relatively high reflection $|S_{22}|$ is not a concern, since propagation is from port~$1$ to port~$2$, as long as good matching is achieved at port $1$ of loaded C-sections~(see \figref{FIG:LoadedCsec}).

The forward transmission $|S_{21}(2.5\text{ GHz})|\approx 16$~dB, with $|T_{\text{L}}| = 6.7$~dB, corresponding to the enhanced group delay $\tau_{21}(2.5\text{ GHz})\approx3$~ns~[\figref{FIG:Resp_SimSingleGD}], while \mbox{$|S_{21}(2.5\text{ GHz})| \approx 0$~dB}, with $|T_{\text{L}}| = 0$~dB, corresponding to conventional C-section (all pass) with group delay $\tau_{21}(2.5\text{ GHz})\approx1.7$~ns. The loss C-section performance is not simulated here because it is less possible to oscillate than gain C-section, but it will be shown later in the experiment.

\begin{figure}[h!t]
\centering
   \psfragfig*[width=0.5\linewidth, trim={0in 0in 0in 0in}]{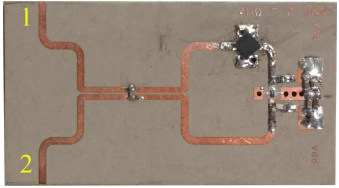}{
   \psfrag{1}[c][c][0.8]{\textcolor{yellow}{1}}
   \psfrag{2}[c][c][0.8]{\textcolor{yellow}{2}}
    }
    \caption{Fabricated loaded C-section phaser.}
    \label{FIG:PhotoLoadedCsec}
\end{figure}

\begin{figure}[h!t]
   \centering
   \psfrag{F}[c][c][0.8]{Frequency (GHz)}
   \subfigure[]{
    \psfragfig*[width=0.95\linewidth, trim={0in 0in 0in 0in}]{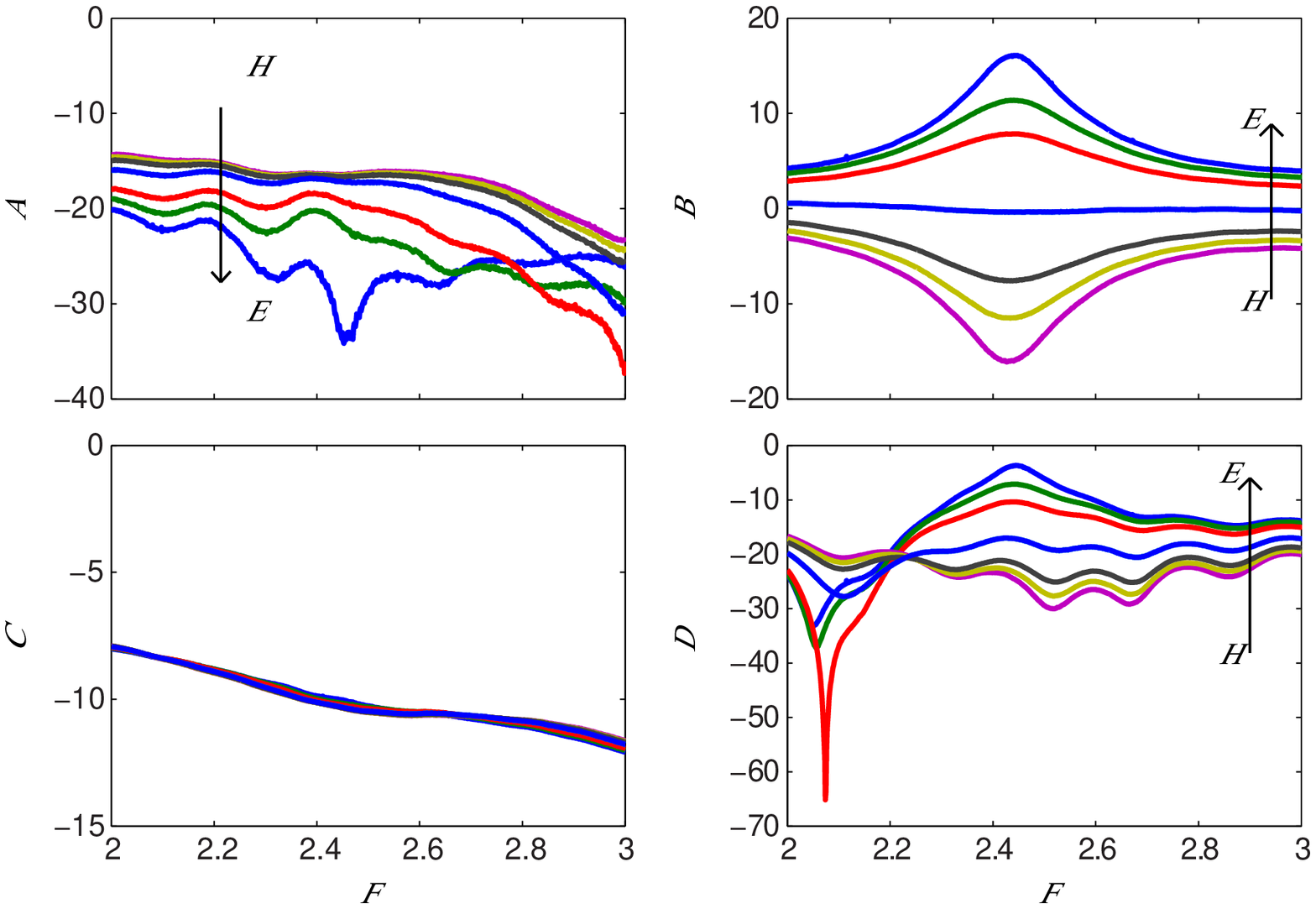}{
    \psfrag{A}[c][c][0.8]{$|S_{11}|$ (dB)}
    \psfrag{B}[c][c][0.8]{$|S_{21}|$ (dB)}
    \psfrag{C}[c][c][0.8]{$|S_{12}|$ (dB)}
    \psfrag{D}[c][c][0.8]{$|S_{22}|$ (dB)}
    \psfrag{E}[r][r][0.8]{$6.7$ dB}
    \psfrag{H}[r][r][0.8]{$-6.7$ dB}
    }
   \label{FIG:Resp_FinishedSingleAMP}
   }

   \subfigure[]{
   \psfragfig*[width=0.6\linewidth, trim={0in 0in 0in 0in}]{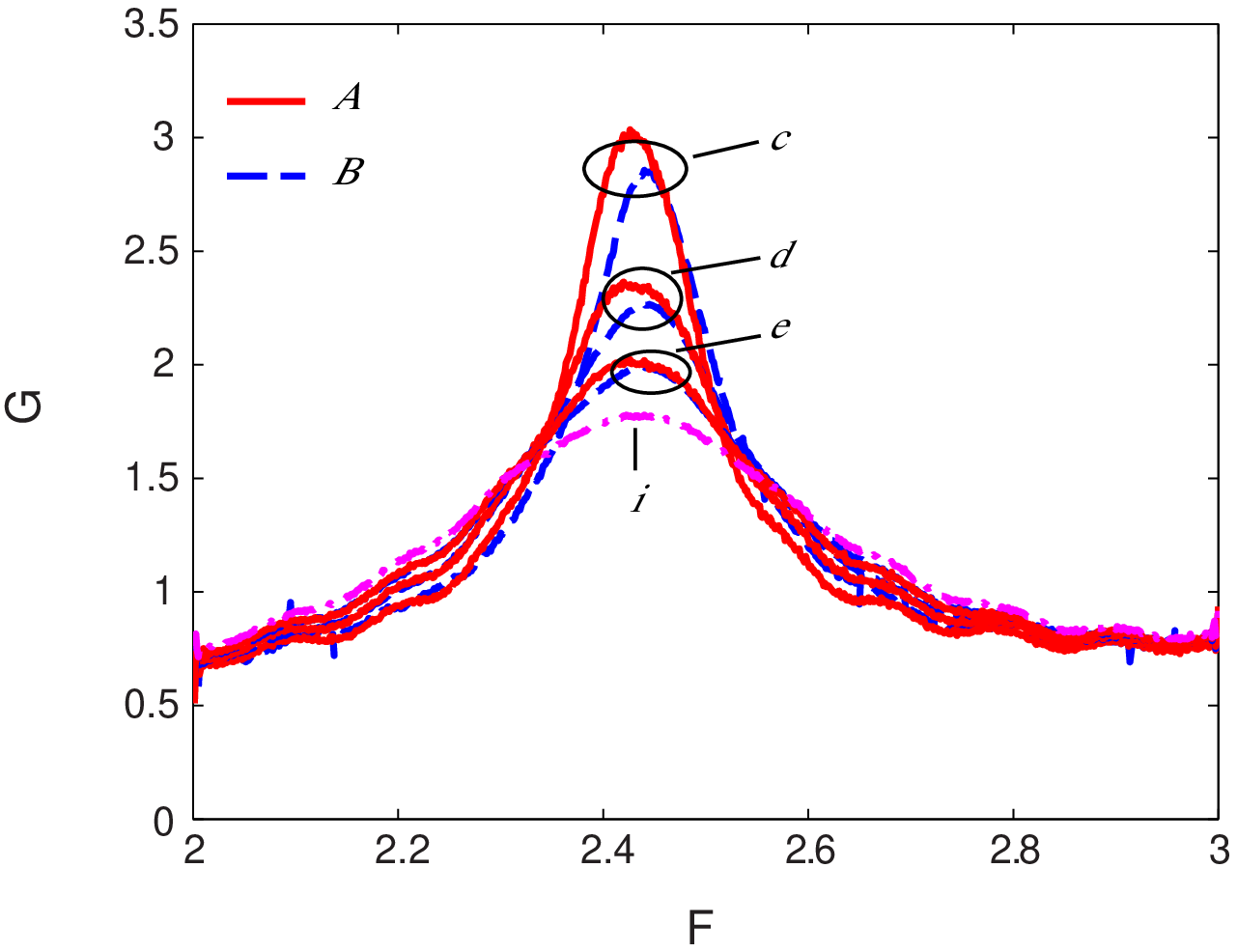}{

    \psfrag{G}[c][c][0.8]{Group delay $\tau_{21}$ (ns)}
    \psfrag{c}[l][l][0.8]{$\pm6.7$ dB}
    \psfrag{d}[l][l][0.8]{$\pm5$ dB}
    \psfrag{e}[l][l][0.8]{$\pm3.7$ dB}
    \psfrag{i}[c][c][0.8]{$0$ dB}
    \psfrag{A}[l][l][0.8]{Loss}
    \psfrag{B}[l][l][0.8]{Gain}
   \label{FIG:Resp_FinishedSingleGD}
   }}
   \caption{Measured (a)~amplitudes and (b)~group delays of the loss or gain C-section in~\figref{FIG:PhotoLoadedCsec}, with varying gains and losses $|T_{\text{L}}|=\{0, \pm3.7, \pm5, \pm6.7\}$~dB.}
   \label{FIG:Resp_FinishedSingle}
\end{figure}

Figure~\ref{FIG:PhotoLoadedCsec} shows the fabricated loaded C-section. The corresponding measured amplitudes and group delays, with varied load loss and gain, are plotted in~\figref{FIG:Resp_FinishedSingle}. Consistently with the analysis presented in Sec.~\ref{SEC:PRINP:PROVE}, the forward transmission amplitudes are symmetric about $0$~dB while the group delays are identical.

\begin{figure}[h!t]
  \centering
  \psfragfig*[width=0.6\linewidth, trim={0in 0in 0in 0in}]{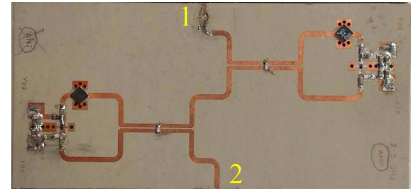}{
  \psfrag{1}[l][l][0.8]{\textcolor{yellow}{1}}
  \psfrag{2}[l][l][0.8]{\textcolor{yellow}{2}}
    }
  \caption{Fabricated equalized loss-gain pair phaser.}
  \label{FIG:Fab_eq_LG_pair}
\end{figure}

\begin{figure}[h!t]
   \centering
   \psfragfig*[width=1\linewidth, trim={0in 0in 0in 0in}]{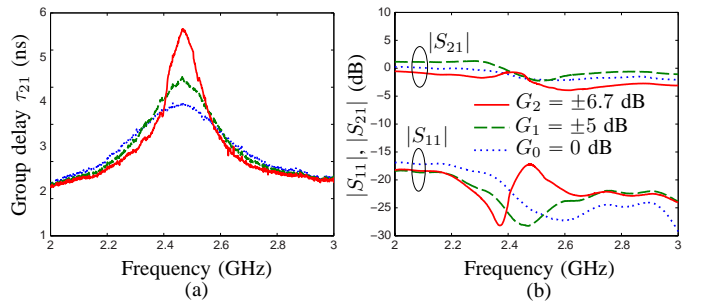}{
    \psfrag{F}[c][c][0.8]{Frequency (GHz)}
    \psfrag{G}[c][c][0.8]{Group delay $\tau_{21}$ (ns)}
    \psfrag{S}[c][c][0.8]{$|S_{11}|$, $|S_{21}|$ (dB)}
    \psfrag{a}[l][l][0.8]{$G_2 = \pm{6.7}$ dB}
    \psfrag{b}[l][l][0.8]{$G_1 = \pm{5}$ dB}
    \psfrag{c}[l][l][0.8]{$G_0 = 0$ dB}
    \psfrag{d}[c][c][0.8]{(a)}
    \psfrag{e}[c][c][0.8]{(b)}
    \psfrag{T}[l][l][0.8]{$|S_{21}|$}
    \psfrag{R}[l][l][0.8]{$|S_{11}|$}
    }
   \caption{Measured (a)~group delays and (b)~amplitudes  of the loss-gain pair phaser in~\figref{FIG:Fab_eq_LG_pair}.}
   \label{FIG:Resp_FinishedEqualized}
\end{figure}

Figure~\ref{FIG:Fab_eq_LG_pair} shows the fabricated equalized loss-gain pair phaser. The corresponding measured amplitudes and group delays, with varied load loss and gain, are plotted in~\figref{FIG:Resp_FinishedEqualized}. Consistently with the analysis presented in Sec.~\ref{SEC:CombinedCell}, the combined pair transmission amplitudes are nearly flat while combined pair group delays are twice of the loss or gain C-section~[compare to \figref{FIG:Resp_FinishedSingleGD}].

\section{Reconfigurable Cascaded Loss-Gain Pair Phaser}\label{SEC:SYN}

\subsection{Constitutive Loss-Gain Pairs}

\begin{figure}[h!t]
  \centering
  \psfragfig*[width=0.7\linewidth, trim={0in 0in 0in 0in}]{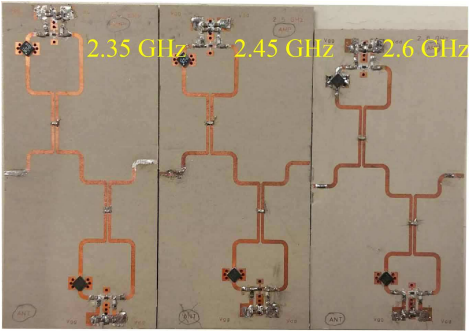}{
  \psfrag{1}[l][l][0.8]{\textcolor{yellow}{$2.35$ GHz}}
  \psfrag{2}[l][l][0.8]{\textcolor{yellow}{$2.45$ GHz}}
  \psfrag{3}[l][l][0.8]{\textcolor{yellow}{$2.6$ GHz}}
    }
  \caption{Fabricated loss-gain equalized pairs, tuned at $2.35$, $2.45$ and $2.6$~GHz, respectively, by varying the length of the loads. The incorporated couplers are all identical.}
  \label{FIG:SynthesizedHardware}
\end{figure}

Cascading loss-gain equalized pairs tuned at different resonance frequencies allows to synthesize reconfigure group delay responses in real time with all-pass amplitude response over a certain bandwidth. To demonstrate this, we fabricated three loss-gain equalized pairs, shown in~\figref{FIG:SynthesizedHardware}, tuned at $\omega_\text{p1}=2.35$, $\omega_\text{p2}=2.45$ and $\omega_\text{p3}=2.6$ GHz, respectively.

\subsection{Experimental Results}\label{SEC:DEMO}

\begin{figure}[h!t]
\centering
\subfigure[]{\label{FIG:PhotoSetup}
   \psfragfig*[width=1\linewidth, trim={0in 0in 0in 0in}]{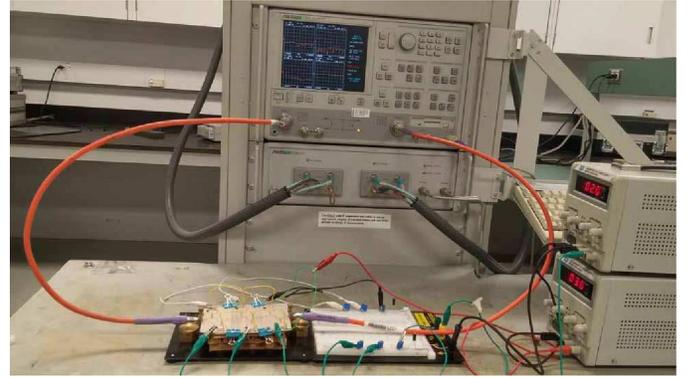}{
    }}
\subfigure[]{\label{FIG:PhotoDUT}
   \psfragfig*[width=1\linewidth, trim={0in 0in 0in 0in}]{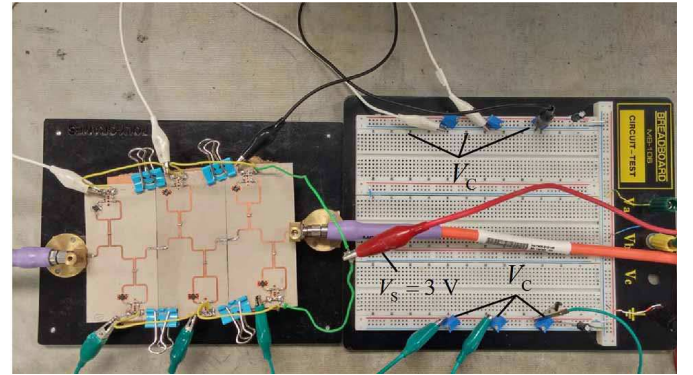}{
   \psfrag{A}[l][l][1]{$V_\text{c}$}
   \psfrag{B}[l][l][1]{$V_\text{c}$}
   \psfrag{C}[l][l][1]{$V_\text{s}=3$ V}
    }}
    \label{FIG:PhotoExpriment}
    \caption{(a)~Experimental setup and (b)~device under test, where a 3-section loss-gain reconfigurable phaser is formed by cascading three loss-gain equalized pairs.}
\end{figure}

The experimental setup and complete device under test are shown in~\figref{FIG:PhotoSetup} and~\figref{FIG:PhotoDUT}, respectively. The 3-section reconfigurable phaser is formed by cascading the three loss-gain equalized pairs, which were clipped on a copper plate placed underneath the pairs. The scattering matrix and group delay were measured on a vector network analyzer. The two power supplies provide a source voltage of $V_\text{s}=3$ V and a control voltage of $V_\text{c}=2$ V, respectively.

\begin{figure}[h!t]
   \centering
   \psfragfig*[width=1\linewidth, trim={0in 0in 0in 0in}]{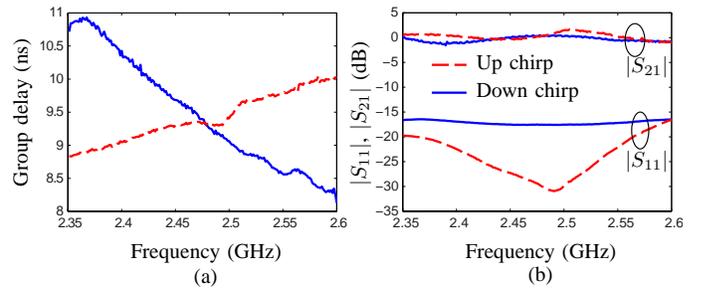}{
    \psfrag{F}[c][c][0.8]{Frequency (GHz)}
    \psfrag{A}[c][c][0.8]{Group delay (ns)}
    \psfrag{B}[c][c][0.8]{$|S_{11}|$, $|S_{21}|$ (dB)}
    \psfrag{c}[l][l][0.8]{Up chirp }
    \psfrag{d}[l][l][0.8]{Down chirp}
    \psfrag{a}[c][c][0.8]{(a)}
    \psfrag{b}[c][c][0.8]{(b)}
    \psfrag{R}[l][l][0.8]{$|S_{11}|$}
    \psfrag{T}[l][l][0.8]{$|S_{21}|$}
    }
   \caption{(a)~Group delays and (b)~Transmission and reflection amplitudes of the cascade 3-section reconfigurable phaser.}
   \label{FIG:Resp_FinishedCascade}
\end{figure}

In Fig.~\ref{FIG:Resp_FinishedCascade} the control voltages ($V_\text{c}$) of the~3 loss-gain pairs (6~voltages in all) are tuned to produce up-chirp and down-chirp linear group delay responses~\cite{Jour:2013_MwMag_Caloz}. This demonstrates the central point of the paper: the group delay response of the phaser can be reconfigured in real time with essentially all-pass transmission. The reconfigurability shown here, between positive and negative chirp responses, is naturally only an illustrative choice, other group delay responses being achievable with this phaser.

\section{Conclusion}\label{SEC:CONCLU}

A loss-gain equalized reconfigurable phaser has been proposed, analyzed and demonstrated. Experimental results have confirmed that such a device provides real-time group delay reconfigurability while exhibiting an all-pass response. It will enable radio analog signal processing (R-ASP) systems requiring dynamic adaptability, as for instance dispersion code multiple access~(DCMA).

\bibliographystyle{IEEEtran}
\bibliography{IEEEabrv,REF_MTT}

\end{document}